\documentclass[aps,prb,twocolumn,superscriptaddress]{revtex4}

\usepackage{graphicx}
\usepackage{amsmath}

\begin{document}
\title{Coherent spin manipulation in an exchange-only qubit}

\title{Coherent spin manipulation in an exchange-only qubit}
\author{E. A. Laird}
\altaffiliation{Present address: Kavli Institute of Nanoscience, Delft University of Technology, 2600 GA Delft, The Netherlands}
\affiliation{Department of Physics, Harvard University, Cambridge, Massachusetts 02138, USA}
\author{J. M. Taylor}
\affiliation{National Institute of Standards and Technology, Gaithersburg, Maryland 20899, USA}
\author{D. P. DiVincenzo}
\affiliation{IBM T. J.  Watson Research Center, Yorktown Heights, New York 10598, USA}
\author{C. M. Marcus}
\affiliation{Department of Physics, Harvard University, Cambridge, Massachusetts 02138, USA}
\author{M. P. Hanson}
\affiliation{Materials Department, University of California, Santa Barbara, California 93106, USA}
\author{A. C. Gossard}
\affiliation{Materials Department, University of California, Santa Barbara, California 93106, USA}

\begin{abstract}

Initialization, manipulation, and measurement of a three-spin qubit are demonstrated using a few-electron triple quantum dot, where all operations can be driven by tuning the nearest-neighbor exchange interaction. Multiplexed reflectometry, applied to two nearby charge sensors, allows for qubit readout. Decoherence is found to be consistent with predictions based on gate voltage noise with a uniform power spectrum. The theory of the exchange-only qubit is developed and it is shown that initialization of only two spins suffices for operation.  Requirements for full multi-qubit control using only exchange and electrostatic interactions are outlined.

\end{abstract}

\maketitle

\section{Introduction}

Electron spins confined in quantum dots are an attractive basis for quantum computing because of their long coherence times and potential for scaling~\cite{Loss:1998p271, DiVincenzo:2000p239, Taylor:2005p178}. In the simplest proposal~\cite{Loss:1998p271}, single spins form the logical basis, with single-qubit operations via spin resonance~\cite{Koppens:2006p7}. An alternative scheme, with logical basis formed from singlet and triplet states of two spins~\cite{Levy:2002p281, Petta:2005p188, Taylor:2005p178} requires inhomogeneous static magnetic field for full single-qubit control~\cite{Foletti:2009p781}. Using three spins to represent each qubit removes the need for an inhomogeneous field; exchange interactions between adjacent spins suffice for all one- and two-qubit operations \cite{DiVincenzo:2000p239, Kempe:2001p873}. In this paper, we experimentally demonstrate coherent spin manipulation in a three-spin qubit defined in a triple quantum dot. Initialization, spin manipulation, and measurement of the qubit state using multiplexed reflectometry\cite{Stevenson:2002p768,Reilly:2007p758} are demonstrated. The gate noise is estimated based on decoherence rates.

The interactions of three spins have been explored experimentally~\cite{Step:1994p767} and theoretically~\cite{Buchachenko:2002p81} in the context of physical chemistry, where the recombination of two radicals, originally in an unreactive triplet state, can be catalyzed by exchange with a third spin. Few-electron triple quantum dots~\cite{Gaudreau:2006p62,Schroer:2007p354, Gaudreau:2009p833} have been used to realize charge reconfigurations corresponding to the elementary operations of quantum cellular automata~\cite{Toth:2001p766}, although tunable spin interactions have not yet been demonstrated~\cite{Gaudreau:2008p718}.

\section{Device and measurement scheme}

\begin{figure*}
\center
\includegraphics[width=5.5 in]{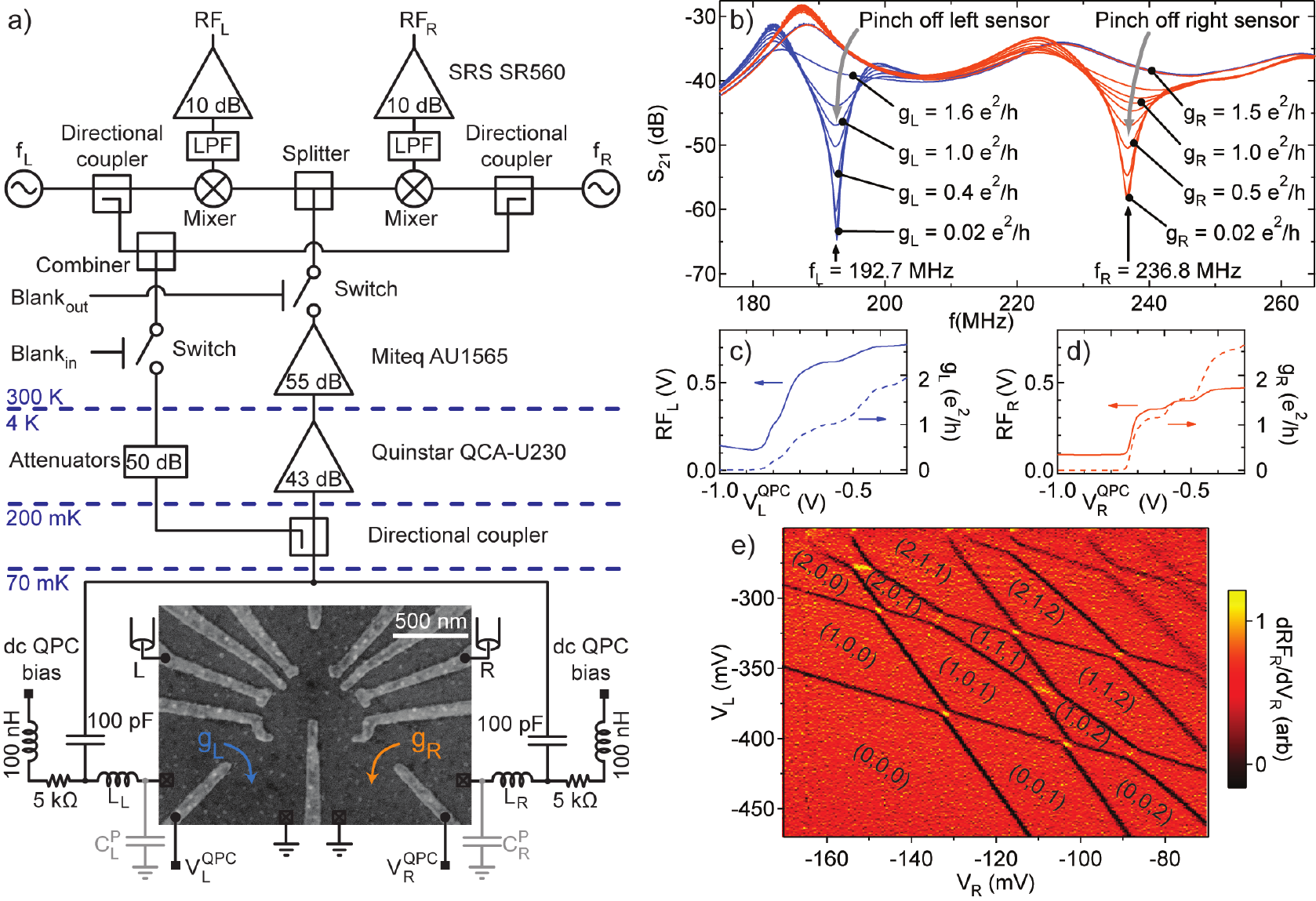}
\caption{\footnotesize{(a) Device and measurement circuit. Patterned topgates define three quantum dots and QPC charge sensors on left and right; voltages applied to gates L and R control the energy levels of the device, while voltages~$V_\mathrm{L}^{QPC}$ and $V_\mathrm{R}^{QPC}$ tune QPC conductances $g_\mathrm{L}$ and $g_\mathrm{R}$. The QPCs are incorporated into resonant tank circuits comprising chip inductors $L_\mathrm{L}$ and $L_\mathrm{R}$ combined with parasitic capacitances $C_\mathrm{L}^P$ and $C_\mathrm{R}^P$; bias tees allow the QPCs to be measured both at DC and via RF reflectometry. An RF carrier, generated by combining signals at resonant frequencies~$f_\mathrm{L}$ and~$f_\mathrm{R}$, is applied to the device via a directional coupler; the reflected signal, after amplification, is demodulated by mixing with the original carrier frequencies to yield voltages $V_\mathrm{L}^\mathrm{RF}$ and $V_\mathrm{R}^\mathrm{RF}$ sensitive predominantly to left and right QPCs respectively. (b) Reflectance~$S_{21}(f)$ as a function of frequency, $f$, of the combined tank circuits, measured with a network analyzer, as the QPCs are pinched off, showing separate resonances corresponding to left and right. (c) and (d), QPC pinchoff measured simultaneously in reflectometry and DC conductance. (e) Reflectometry signal for the right sensor measured as a function of $V_\mathrm{L}$ and $V_\mathrm{R}$, showing steps corresponding to charge transitions. Electron configurations for each gate setting are indicated.}}
\end{figure*}

We first demonstrate how our device~(Fig.~1(a)) can be operated in the three-electron regime, then discuss coherent manipulation of the three-spin system. The device was fabricated by patterning Ti/Au topgates on a GaAs/AlGaAs heterostructure incorporating a two-dimensional electron gas 110~nm beneath the surface. Depletion gate voltages create a triple quantum dot together with a pair of charge sensing quantum point contacts~(QPCs)~\cite{FIELD:1993p238}. Gates~L and~R are connected to coaxial lines allowing rapid voltage pulses to be applied. The device was measured at 150~mK electron temperature in a dilution refrigerator equipped with an in-plane magnetic field.

A frequency-multiplexed radio-frequency~(RF) reflectometry circuit~\cite{Stevenson:2002p768, Reilly:2007p758} allowed both QPCs to be measured independently with MHz bandwidth~(Fig.~1(a)). Parallel resonant tank circuits incorporating left and right QPCs were formed from nearby inductors $L_\mathrm{L}=910$~nH and $L_\mathrm{R}=$750~nH together with the parasitic capacitances $C^\mathrm{P}_\mathrm{L}$ and $C^\mathrm{P}_\mathrm{R}$ of the bond wires. Bias tees coupled to each tank circuit allowed the DC conductances $g_\mathrm{L}$, $g_\mathrm{R}$ of left and right QPCs to be measured simultaneously with the reflectance of the RF circuit.  As each QPC was pinched off, a separate dip developed in the reflected signal at corresponding resonant frequency~$f_\mathrm{L,R}\approx (2\pi)^{-1}(L_\mathrm{L,R}C^\mathrm{P}_\mathrm{L,R})^{-1/2}$~(Fig.~1(b)). To monitor the charge sensors, two carrier frequencies $f_\mathrm{L}$ and~$f_\mathrm{R}$ were applied to the single coaxial line driving both resonant circuits~(Fig.~1(a)). The reflected signal was amplified using both cryogenic and room temperature amplifiers, then demodulated by mixing with local oscillators and low-pass filtered to yield voltages $V^\mathrm{RF}_\mathrm{L}$ and $V_\mathrm{R}^\mathrm{RF}$ sensitive predominantly to $g_\mathrm{L}$ and $g_\mathrm{R}$~(Figs.~1(c) and (d)). To suppress back-action and reduce pulse coupling into the readout circuit, the RF carrier was blanked on both signal and return paths except during the readout pulse configuration; no RF was applied to the readout circuit during spin initialization and manipulation.

With $g_\mathrm{R}$ tuned to the point of maximum charge sensitivity~$g_\mathrm{R}\sim0.4e^2/h$, the configuration of the triple dot was monitored~\cite{Reilly:2007p758} via $V_\mathrm{R}^\mathrm{RF}$. Sweeping voltages $V_\mathrm{L}$ and $V_\mathrm{R}$ on gates L and R, the charge stability diagram of the triple dot was mapped out, as shown in Fig.~1(e)). Dark transition lines are seen to run with three different slopes, corresponding to electrons added to each of the three dots~\cite{Gaudreau:2006p62, Schroer:2007p354}. For the most negative voltages, transitions are no longer seen, indicating that the device has been completely emptied. This allows absolute electron occupancies of the three dots to be assigned to each region of the diagram.

\section{Exchange-only qubit operation}
\begin{figure*}
\center \label{fig:triple2}
\includegraphics[width=5.5 in]{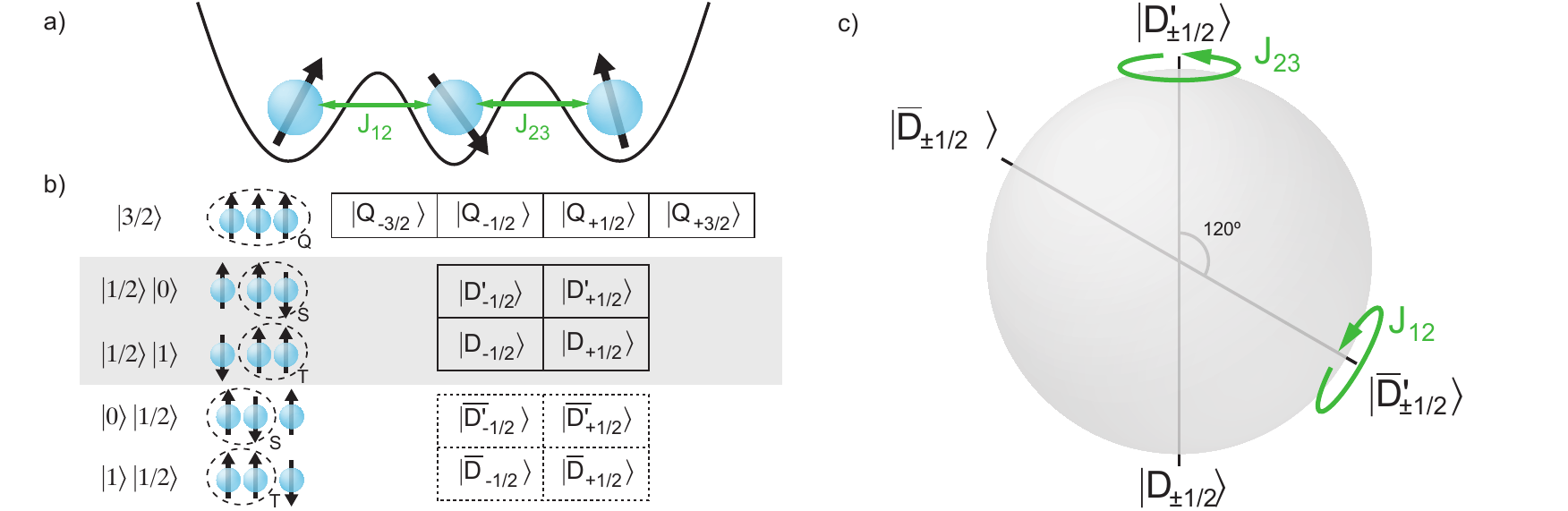}
\caption{\footnotesize{An exchange-only qubit. (a) Electron spins in three adjacent quantum dots are coupled by nearest-neighbour exchange. (b), The eight states of the system can be divided into a quadruplet, $Q$, and two doublets, $D'$ and $D$, distinguished by the multiplicity (singlet or triplet) of the rightmost pair of spins. An alternative choice, denoted $\overline{D}$ and $\overline{D'}$, distinguishes the doublets according to the multiplicity of the leftmost spin pair (dashed boxes). (c), Choosing an element from each doublet as the qubit basis (highlighted in (b)), arbitrary unitary transformations are equivalent to rotations on the Bloch sphere shown, where doublet states $|D'_{\pm1/2}\rangle$ and $|D_{\pm1/2}\rangle$ correspond to north and south poles and states $|\overline{D'}_{\pm1/2}\rangle$ and $|\overline{D}_{\pm1/2}\rangle$ to poles of an axis tilted by $120^\circ$.  Exchange between middle and right dots drives rotations about the $D-D'$ axis, while exchange between left and middle dots drives rotations about the $\overline{D}-\overline{D'}$ axis. In combination, any rotation can be accomplished.}}
\end{figure*}

\subsection{The qubit subspace}

We work in the subspace of three electrons restricted to occupancies of at most two electrons per dot. To see how exchange can drive arbitrary qubit operations, consider three spins coupled by nearest-neighbour exchange strengths $J_{12}$ and~$J_{23}$~(Fig.~2(a))~\cite{DiVincenzo:2000p239}. The eight spin states can be classified by both overall multiplicity and multiplicity of the rightmost spin pair, and comprise a quadruplet, $|Q_{S_z}\rangle$, and two doublets, $|D'_{S_z}\rangle$ and $|D_{S_z}\rangle$, where $S_z$ denotes the $z$-component  of total spin and takes values~$S_z=\pm 1/2$ or $\pm 3/2$ for the quadruplet and~$S_z=\pm 1/2$ for the doublets~(Fig.~2(b))~\cite{Knill:2000p871,Yang:2001p872, Buchachenko:2002p81}. Whereas for $|D'_{S_z}\rangle$ states, the rightmost pair of spins forms a singlet, for $|D_{S_z}\rangle$ states, the rightmost pair forms a mixture of triplet states (see Appendix \ref{appendixB}). Alternatively, the doublets can be classified according to the multiplicity of the leftmost pair: States $|\overline{D'}_{S_z}\rangle$ correspond to singlets on the left whereas states $|\overline{D}_{S_z}\rangle$ correspond to triplet states.

The logical basis is formed from two states with equal~$S_z$, one taken from each doublet $|D'_{S_z}\rangle$ and $|D_{S_z}\rangle$. That is, we define the logical qubit states $|0\rangle$ and $|1\rangle$ as $|0\rangle=|D_{\pm1/2}\rangle$ and  $|1\rangle=|D'_{\pm1/2}\rangle$~(Fig.~1)~\cite{DiVincenzo:2000p239}. A valid qubit can be formed from either $S_z=+1/2$ or $S_z=-1/2$ doublet components, or any mixture of the two; it is therefore necessary to prepare and read out only two of the three spins in order to implement full single-qubit operation. We do not discuss further the spin-3/2 subspace, as we start only from states with spin 1/2 and do not otherwise change the total spin.

States of the qubit correspond to points on the Bloch sphere shown in Fig.~2(c). Exchange~$J_{23}$ between the rightmost spin pair drives qubit rotations about the vertical axis, exchange $J_{12}$ between the leftmost pair drives rotations about an axis tilted by $120^\circ$ and defined by doublets $|\overline{D'}_{S_z}\rangle$ and~$|\overline{D}_{S_z}\rangle$. Arbitrary single-qubit operations can be achieved by concatenating up to four exchange pulses~\cite{DiVincenzo:2000p239}.
\begin{figure}[b]
\center \label{fig:triple3}
\includegraphics[width=3 in]{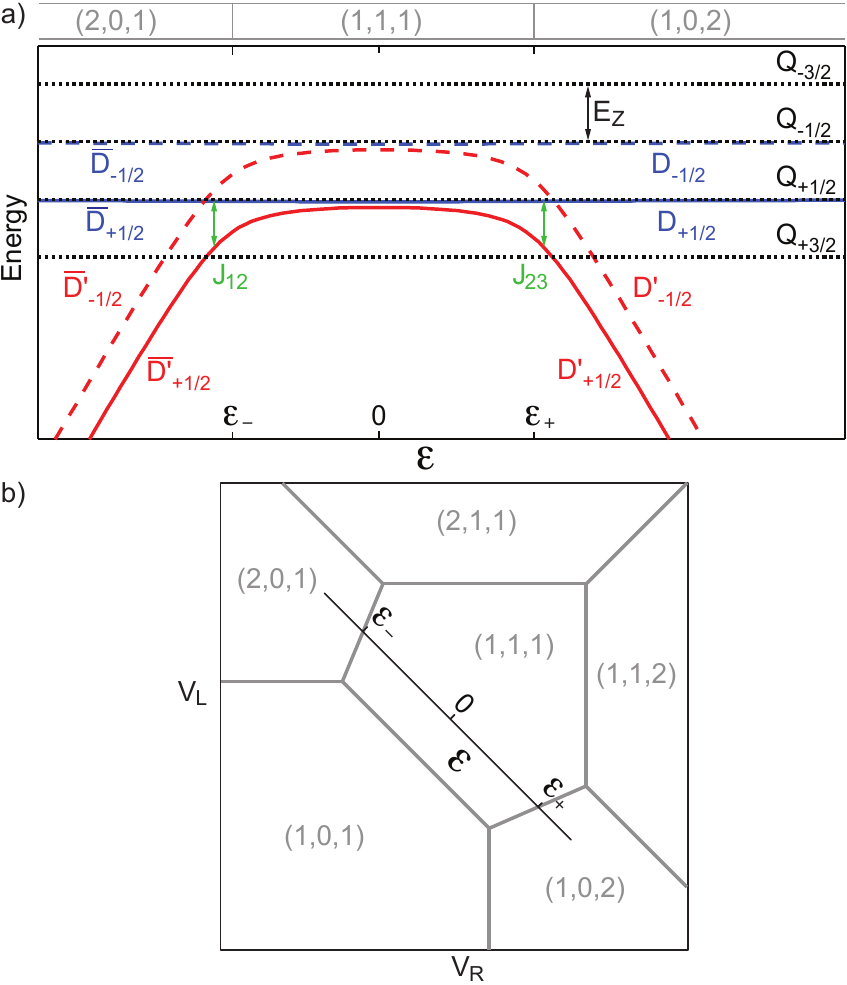}
\caption{\footnotesize{(a) Three-electron energy levels as a function of detuning $\epsilon$, showing Zeeman and exchange splitting (see Appendix \ref{appendixA} for details of calculation). The case where left and right inter-dot tunnel couplings are equal is plotted; the case of strong asymmetry, corresponding to the experiment, is discussed in Appendix~\ref{appendixC}. Near zero detuning the device is configured in (1,1,1) with negligible exchange; increasing (decreasing) $\epsilon$  lowers the energy of the $D'$ ($\overline{D'}$) doublet by exchange $J_{23}$ ($J_{12}$). For $\epsilon>\epsilon_+$ ($\epsilon<\epsilon_-$), states in doublet $D'$ ($\overline{D'}$) correspond to a predominant~(1,0,2) ((2,0,1)) configuration. (b) Ground-state configuration of a triple dot as a function of gate voltages $V_\mathrm{L}$ and $V_\mathrm{R}$ coupled to left and right dots~\cite{Schroer:2007p354}. The detuning axis is shown.}}
\end{figure}

\subsection{Tuning the exchange interaction}

The device energy levels are tuned with an external magnetic field $B$ and by using gate voltages to adjust the energies of different charge configurations $(N_\mathrm{L}, N_{\rm M}, N_\mathrm{R})$, where $N_\mathrm{L}$, $N_{\rm M}$ and $N_\mathrm{R}$ denote electron occupancies of left, middle and right dots respectively (see Appendix~\ref{appendixA}). Defining detuning $\epsilon$ as the energy difference between (2,0,1) and (1,0,2) configurations (in units of gate voltage), three regimes are accessible~(Fig.~3(a)). Neglecting hyperfine coupling, the energy levels are set mainly by the exchange interaction and the Zeeman energy $E_Z=g \mu_B B$, where $g$ is the electron $g$-factor and $\mu_B$ is the Bohr magneton. Near $\epsilon=0$, the device is in the (1,1,1) configuration with negligible exchange. As $\epsilon$ is increased, hybridization between (1,1,1) and (1,0,2) configurations lowers the energy of $|D'_{S_z}\rangle$ states, until for $\epsilon>\epsilon_+$, the ground state configuration becomes predominantly (1,0,2). An exchange splitting $J_{23}$  for $\epsilon>0$ prevents occupation of the (1,0,2) configuration with $|Q_{S_z}\rangle$ and $|D_{S_z}\rangle$ spin states and enforces Pauli exclusion in the rightmost dot.  Similarly, with decreasing $\epsilon$ the energy of $|\overline{D}_{S_z}\rangle$ states is lowered by an amount $J_{12}$, and below $\epsilon=\epsilon_-$ the ground state configuration becomes predominantly (2,0,1). The various configurations are accessed by tuning gate voltages~$V_\mathrm{L}$ and~$V_\mathrm{R}$ coupled predominantly to  left and right dots respectively. The lowest-energy configurations of three capacitively coupled dots are modeled in Fig.~3(b), which also illustrates the detuning axis in gate space.

\begin{figure}
\center \label{fig:triple4}
\includegraphics[width=3.2 in]{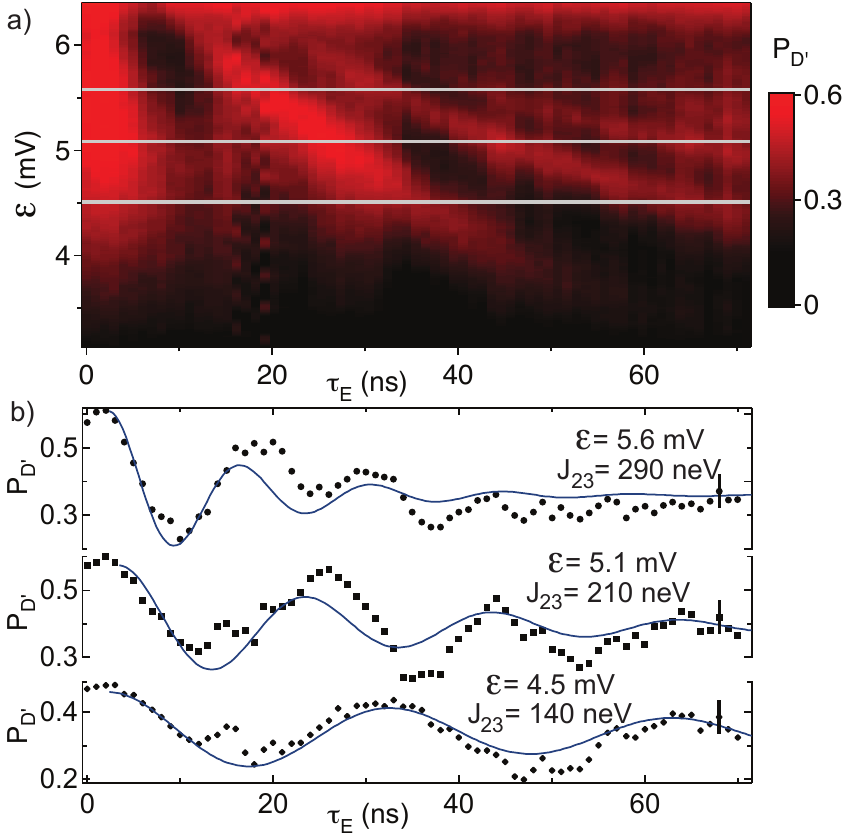}
\caption{\footnotesize{Coherent spin exchange. (a) Probability
 $P_{D'}$ to return to the initial $|D'_{S_z}\rangle$ state following an exchange pulse sequence, measured as a function of $\epsilon$ during the exchange pulse and pulse duration $\tau_{\rm E}$. Dark and bright regions respectively indicate odd and even numbers of complete spin exchanges. (b), Points: Measured $P_{D'}$ as a function of $\tau_{\rm E}$ for values of $\epsilon$ indicated by horizontal lines in (a). Lines: Fits to exponentially damped phase-shifted cosines, corresponding to coherent rotations dephased by electric fields with a white noise spectrum (see text). The fitted exchange $J_{23}(\epsilon)$ for each curve is shown.}}
\end{figure}

\subsection{Coherent spin manipulation}
Repeated spin state initialization, coherent manipulation, and readout uses the following cycle of voltage pulses~\cite{Petta:2005p188} on gates L and R to rapidly tune $\epsilon$: Beginning at~$\epsilon>\epsilon_+$ configures the device in (1,0,2) where tunneling to the leads initializes the qubit within the doublet~$|D'_{S_z}\rangle$. The detuning is then decreased to $\epsilon \sim 0$ over~1~$\mu$s, configuring the device in (1,1,1). Because this ramp time is adiabatic compared to the characteristic hyperfine interaction strength, the spin system enters a ground state defined by the instantaneous nuclear configuration, for example $|\hspace{-0.1 cm}\uparrow \downarrow \uparrow \rangle$~\cite{Petta:2005p188, Taylor:2007p719}.  Pulsing the detuning close to $\epsilon_+$, where $J_{23}$ is large, for a time $\tau_\mathrm{E}$ leads to coherent exchange of spins between the right-hand dots. Finally, the detuning is ramped back to its original value $\epsilon>\epsilon_+$. The charge configuration is now determined by the outcome of the exchange pulse: Whereas the hyperfine ground state reenters the $|D'_{S_z}\rangle$ doublet in the (1,0,2) configuration, a swapped state such as $|\hspace{-0.1 cm}\uparrow \uparrow \downarrow \rangle$ evolves into a superposition of $|D_{S_z}\rangle$ and $|Q_{\pm1/2}\rangle$ states, causing the device to remain in (1,1,1).  At the end of this final ramp, the carrier is unblanked for readout of the charge sensor. Waiting another $\sim 5$ $\mu$s reinitializes the spin state and the cycle begins again. 

Averaged over $\sim 1000$ cycles, the resulting voltage~$V_\mathrm{R}^\mathrm{RF}$ is converted to a spin state probability by calibrating it against ~$V_\mathrm{R}^\mathrm{RF}$ values corresponding to (1,1,1) and (1,0,2) configurations. The probability $P_{D'}$ to return to the initial spin state is shown in Fig.~4(a) as a function of $\tau_{\rm E}$ and $\epsilon$ during the exchange pulse.  As a function of $\tau_{\rm E}$,  $P_{D'}$ oscillates showing coherent rotation between spin states, and the oscillation frequency, set by $J_{23}(\epsilon)$, increases with $\epsilon$ as expected from Fig.~3(a). The measured $P_{D'}(\tau_{\rm E})$ is fitted for three values of $\epsilon$ with an exponentially damped cosine, corresponding to dephasing by electric fields with a white noise spectrum~\cite{Petta:2005p188, Taylor:2007p719} (Fig.~4(b)). The extracted $J_{23}(\epsilon)$ depends exponentially on $\epsilon$, similar to observations at comparable exchange strength in a double dot~\cite{Foletti:2009p781}, but inconsistent with the power-law dependence found at more negative detunings~\cite{Laird:2006p28}. 

Experimental $P_{D'}(\tau_{\rm E})$ values in Fig.~4(b) are fit to an exponentially damped cosine form, $P_{D'}(\tau_{\rm E}) = A e^{-\alpha \tau_{\rm E}}\cos(J_{23} \tau_{\rm E} / h + \phi) + B$, where $\alpha$ is a damping coefficient reflecting decoherence presumably attributable to gate voltage noise \cite{Taylor:2007p719}. This form is appropriate for a white noise spectrum, and was chosen over alternative forms (with higher powers of $\tau_{\rm E}$ appearing in the exponent) by the quality of fit, judged by eye. $A$, $B$, and $\phi$ are phenomenological amplitude, offset, and phase parameters.  A value for the voltage noise spectral density of detuning, $\sigma_\epsilon  =  \hbar \alpha^{1/2}/(dJ_{23}/d\epsilon)= 27\pm5 \,\,\mathrm{nV}/\sqrt{\mathrm{Hz}}$, was obtained from a fit to the top data set in Fig.~4(b), using an independently measured value $dJ_{23}/d\epsilon$. The lower two curves use the same value of $\sigma_\epsilon$ with independently measured values of $dJ_{23}/d\epsilon$, and show equally good agreement with the data. The origin of this surprisingly large voltage noise, accounting for the observed rapid decoherence, is presently unknown.  Reduced contrast ($A <1$) can be attributed to pulse imperfections~\cite{Foletti:2009p781}, which also cause a small phase shift. Similar data for $J_{12}$ could not be obtained in this device due to weak tunnel coupling between left and middle dots (see Appendix~\ref{appendixC}).

In summary, we have fabricated a three-electron spin qubit and demonstrated initialization, coherent spin manipulation using pulsed-gate control of exchange, and state readout. These operations do not yet constitute full qubit control, however. For that, pulsed operation of both $J_{12}$ and $J_{23}$ is needed. Furthermore, to complete a universal set of gates, two-qubit operations will also be needed. That could be done with nearest neighbor exchange coupling of two three-spin qubits, as described in  Refs.~\onlinecite{DiVincenzo:2000p239, Kawano:2004p840}, which require that the third spin be initialized into a known state. Capacitive coupling of two three-spin qubits can also form a two qubit gate, and does not require initializing the third spin ~\cite{Taylor:2005p178}. Those tasks, along with reducing electrical noise to improve coherence, remain for future work.

\begin{acknowledgments}
We acknowledge C. Barthel and D.~J. Reilly for discussions. This work was supported by the Department of Defense, IARPA/ARO, the National Science Foundation, and Harvard University.
\end{acknowledgments}

\appendix
\section{Energy levels of three exchange-coupled spins}
\label{appendixA}
In this Appendix we present the states and energy levels of three electron spins as shown in Fig.~1(a), coupled by nearest-neighbour exchange  and subject to a magnetic field. The Hamiltonian is\cite{Buchachenko:2002p81}:
\begin{multline}
\label{eq:hamiltonian}
H=J_{12}\left({\bf S}_1\cdot {\bf S}_2-\frac{1}{4}\right)+J_{23}\left({\bf S}_2\cdot {\bf S}_3-\frac{1}{4}\right)\\
 -E_\mathrm{Z}(S_1^z+S_2^z+S_3^z),
 \end{multline}
where the spins are denoted $\mathbf{S}_1$, $\mathbf{S}_2$, $\mathbf{S}_3$, the magnetic field is along the $z$-axis, and units are chosen so that Planck's constant is $\hbar=1$.

The eight spin eigenstates of the Hamiltonian~(\ref{eq:hamiltonian}) form a quadruplet $Q$ and high- and low-energy doublets $\Delta, \Delta'$:
\vspace{2cm}
\begin{widetext}
\begin{eqnarray}
|Q_{+3/2}\rangle &=& |\uparrow\uparrow\uparrow\rangle \\
|Q_{+1/2}\rangle &=& \frac{1}{\sqrt{3}}\left( |\uparrow\uparrow\downarrow\rangle +  |\uparrow\downarrow\uparrow\rangle + |\downarrow\uparrow\uparrow\rangle \right) \\
|Q_{-1/2}\rangle &=& \frac{1}{\sqrt{3}}\left( |\downarrow\downarrow\uparrow\rangle +  |\downarrow\uparrow\downarrow\rangle + |\uparrow\downarrow\downarrow\rangle \right) \\
|Q_{-3/2}\rangle &=& |\downarrow\downarrow\downarrow\rangle \\
|\Delta_{+1/2}\rangle &=& \frac{1}{\sqrt{4\Omega^2+2\Omega(J_{12}-2J_{23})}}
((J_{12}-J_{23}+\Omega)|\uparrow\uparrow\downarrow\rangle
+(J_{23}-\Omega)|\uparrow\downarrow\uparrow\rangle
-J_{12}|\downarrow\uparrow\uparrow\rangle)\\
|\Delta_{-1/2}\rangle &=& \frac{1}{\sqrt{4\Omega^2+2\Omega(J_{12}-2J_{23})}}
((J_{12}-J_{23}+\Omega)|\downarrow\downarrow\uparrow\rangle
+(J_{23}-\Omega)|\downarrow\uparrow\downarrow\rangle
-J_{12}|\uparrow\downarrow\downarrow\rangle)\\
|\Delta'_{+1/2}\rangle &=& \frac{1}{\sqrt{4\Omega^2+2\Omega(2J_{23}-J_{12})}}
((-J_{12}+J_{23}+\Omega)|\uparrow\uparrow\downarrow\rangle
-(J_{23}+\Omega)|\uparrow\downarrow\uparrow\rangle
+J_{12}|\downarrow\uparrow\uparrow\rangle)\\
|\Delta'_{-1/2}\rangle &=& \frac{1}{\sqrt{4\Omega^2+2\Omega(2J_{23}-J_{12})}}
((-J_{12}+J_{23}+\Omega)|\downarrow\downarrow\uparrow\rangle
-(J_{23}+\Omega)|\downarrow\uparrow\downarrow\rangle
+J_{12}|\uparrow\downarrow\downarrow\rangle),
\end{eqnarray}
\end{widetext}
with energies:
\begin{eqnarray}
E_{Q_{S_z}}&=&-E_\mathrm{Z} S_z \\
E_{\Delta_{S_z}}&=&-(J_{12}+J_{23}-\Omega)/2 - E_\mathrm{Z} S_z \\
E_{\Delta'_{S_z}}&=&-(J_{12}+J_{23}+\Omega)/2 -E_\mathrm{Z} S_z ,
\end{eqnarray}
where $\Omega= \sqrt{J_{12}^2+J_{23}^2-J_{12}J_{23}}$. Along the detuning axis of~Fig.~3(b), significant charge hybridization is possible between at most pair of dots, allowing the exchange energies to be approximated by functions appropriate for a double dot~\cite{Taylor:2007p719} $J_{12}(\epsilon)=(\epsilon_--\epsilon)/2+\sqrt{((\epsilon_--\epsilon)/2)^2+4t_\mathrm{L}^2}$ and $J_{23}(\epsilon)=(\epsilon-\epsilon_+)/2+\sqrt{((\epsilon-\epsilon_+)/2)^2+t_\mathrm{R}^2}$, where $t_\mathrm{L}$ and $t_\mathrm{R}$ are the left and right inter-dot tunnel couplings. Figure~3(a) shows the resulting energy levels as a function of $\epsilon$ for a symmetric device ($t_\mathrm{L}$=$t_\mathrm{R}$).

\section{The qubit basis states}
\label{appendixB}
The qubit basis states are the doublet eigenstates of Hamiltonian~(\ref{eq:hamiltonian}) in the limit of vanishing exchange on the left,~$J_{12}/J_{23}\rightarrow 0$. In this limit, corresponding to the right side of~Fig.~3(a), the doublet eigenstates are\cite{Buchachenko:2002p81, DiVincenzo:2000p239}:
\begin{eqnarray}
|\Delta_{+1/2}\rangle &\rightarrow& |D_{+1/2}\rangle = \frac{1}{\sqrt{6}}
(|\uparrow\uparrow\downarrow\rangle
+|\uparrow\downarrow\uparrow\rangle
-2|\downarrow\uparrow\uparrow\rangle)
\rule{1cm}{0pt}\raisetag{35pt}\\
|\Delta_{-1/2}\rangle &\rightarrow&  |D_{-1/2}\rangle = \frac{1}{\sqrt{6}}
(|\downarrow\downarrow\uparrow\rangle
+|\downarrow\uparrow\downarrow\rangle
-2|\uparrow\downarrow\downarrow\rangle)\\
|\Delta'_{+1/2}\rangle &\rightarrow& |D'_{+1/2}\rangle = \frac{1}{\sqrt{2}}
(|\uparrow\uparrow\downarrow\rangle
-|\uparrow\downarrow\uparrow\rangle)
\\
|\Delta'_{-1/2}\rangle &\rightarrow& |D'_{-1/2}\rangle = \frac{1}{\sqrt{2}}
(|\downarrow\downarrow\uparrow\rangle
-|\downarrow\uparrow\downarrow\rangle)
,
\end{eqnarray}
with energies:
\begin{eqnarray}
E_{D_{S_z}}&=&-E_\mathrm{Z}S_z\\
E_{D'_{S_z}}&=&-J_{23}-E_\mathrm{Z}S_z.
\end{eqnarray}
The projection of $|D_{S_z}\rangle$ onto states of the leftmost spins is a mixture of triplet states, whereas the projection of $|D'_{S_z}\rangle$ is a singlet.

Analogously, in the limit of vanishing right-dot exchange $J_{23}/J_{12}\rightarrow 0$ (right side of~Fig.~3(a)), the eigenstates are elements of the $\overline{D}$ and $\overline{D'}$ doublets, related to $D$ and $D'$ states by interchange of left and right spins:
\begin{eqnarray}
|\Delta_{+1/2}\rangle &\rightarrow& -|\overline{D}_{+1/2}\rangle = -\frac{1}{\sqrt{6}}
(|\downarrow\uparrow\uparrow\rangle
+|\uparrow\downarrow\uparrow\rangle
-2|\uparrow\uparrow\downarrow\rangle)\notag\\
\\
|\Delta_{-1/2}\rangle &\rightarrow&  -|\overline{D}_{-1/2}\rangle = -\frac{1}{\sqrt{6}}
(|\uparrow\downarrow\downarrow\rangle
+|\downarrow\uparrow\downarrow\rangle
-2|\downarrow\downarrow\uparrow\rangle)\notag\\
\\
|\Delta'_{+1/2}\rangle &\rightarrow& -|\overline{D'}_{+1/2}\rangle = -\frac{1}{\sqrt{2}}
(|\uparrow\downarrow\uparrow\rangle
-|\downarrow\uparrow\uparrow\rangle)
\\
|\Delta'_{-1/2}\rangle &\rightarrow& -|\overline{D'}_{-1/2}\rangle = -\frac{1}{\sqrt{2}}
(|\downarrow\uparrow\downarrow\rangle
-|\uparrow\downarrow\downarrow\rangle)
.
\end{eqnarray}
The corresponding energies are:
\begin{eqnarray}
E_{\overline{D}_{S_z}}&=&-E_\mathrm{Z}S_z\\
E_{\overline{D}'_{S_z}}&=&-J_{12}-E_\mathrm{Z}S_z.
\end{eqnarray}

\begin{figure}
\center
\label{fig:triple5}
\includegraphics[width=8.6cm]{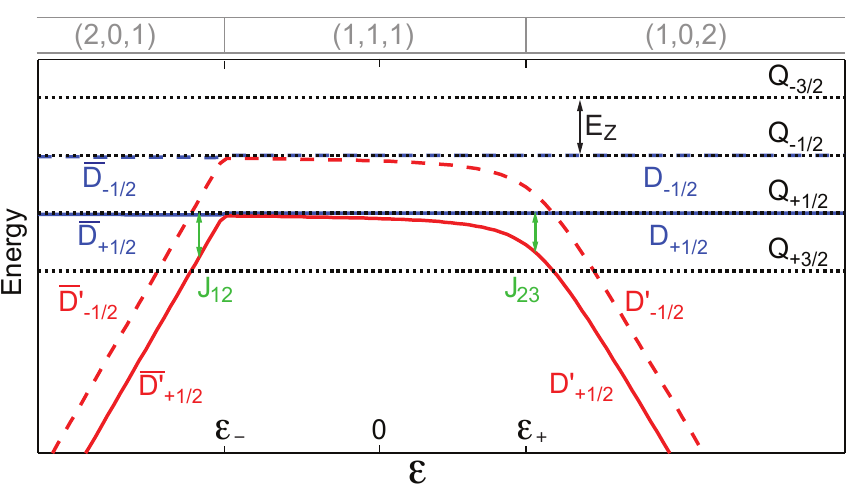}
\caption{\footnotesize{Energy levels of three coupled spins, labelled as in Fig.~3(a), for the case of asymmetric tunnel couplings $t_\mathrm{L} \ll t_\mathrm{R}$. The divergence of the doublet energy levels on the left becomes much sharper, making the effects of $J_{12}$ difficult to observe.}}
\end{figure}
\section{Effect of asymmetric tunnel couplings}
\label{appendixC}

The effect of asymmetric tunnel couplings on the energy levels is shown in~Fig.~5 for the case $t_\mathrm{L}\ll t_\mathrm{R}$. The $\overline{D'}$ levels diverge more abruptly from $\overline{D}$ levels, reducing $J_{12}$ especially for $\epsilon > \epsilon_-$.

A smaller $t_\mathrm{L}$ makes the left-dot exchange harder to observe. The simplest pulse cycle used to study the effects of $J_{12}$ began at $\epsilon>\epsilon_+$, configuring the device in (1,0,2) and initializing the qubit within the doublet $|D'_{S_z}\rangle$. The gate voltages were then rapidly pulsed to $\epsilon<0$ for a time $\tau_S$, during which exchange with the left dot would be expected to drive precession about the $|\overline{D}_{S_z}\rangle - |\overline{D'}_{S_z}\rangle$ axis in Fig.~2(c). For readout, the detuning was returned to $\epsilon>\epsilon_+$, projecting the~$|D'_{S_z}\rangle$ component of the spin state into configuration~(1,0,2) and projecting $|D_{S_z}\rangle$ into (1,1,1). The resulting $P_{D'}(\tau_S)$, measured via reflectometry voltage $V_\mathrm{R}^\mathrm{RF}$, showed no coherent oscillations as a function of $\tau_{S}$; instead a monotonic decay over $\sim10$~ns consistent with hyperfine dephasing~\cite{Petta:2005p188, Taylor:2005p178} was observed. This was true with $\epsilon$ pulsed to either side of $\epsilon_-$ during $\tau_S$.

With energy levels as shown in Fig.~S1, this observation can be explained as follows. For appreciable exchange strength $J_{12}$, $\epsilon$ must be pulsed to $\epsilon<\epsilon_-$ during $\tau_S$. However, precession will only take place if, for the $|\overline{D'}_{S_z}\rangle$ component of the spin state, the configuration (2,0,1) can be accessed. If $t_\mathrm{L}$ is too small, the transition $(1,1,1)\rightarrow(2,0,1)$ cannot occur within $\tau_S$. Instead, the device enters a metastable (1,1,1) configuration (not shown in the level diagrams), where hyperfine coupling incoherently mixes all three multiplets $|D'_{S_z}\rangle$, $|D_{S_z}\rangle$ and $|Q_{S_z}\rangle$.

\bibliography{Triplerefs}

\end{document}